\documentstyle[11pt]{article}

\newcommand{\AmS}{{\protect\the\textfont2
  A\kern-.1667em\lower.5ex\hbox{M}\kern-.125emS}}

\begin{document}
\section*{\center CHOOZ, PALO VERDE, KRASNOYARSK}

\Large \hspace{4cm} L.A. Mikaelyan
\\
\normalsize
\begin{center} 
{\it RRC "Kurchatov Institute", Kurchatov Sq., 1, Moscow-123182, Russia.}
Talk given at TAUP'99, Paris, 6-10 September, 1999. \\
\end{center} 
\righthyphenmin=3

This is a short review of the present $\sim$ 1 km baseline oscillation
experi-ments at nuclear reactors. An idea of a new search for very 
small mixing angle oscillations at Krasnoyarsk is also mentioned. 


\section{INTRODUCTION}

In this report we discuss the following topics:
\begin{itemize}
\item The CHOOZ experiment (France, Italy, Russia, USA): final results 
based on the entire data sample [1]. We refer to the previous
publication [2] for CHOOZ'97 results and the general description 
of the experiment.
\item The Palo Verde experiment (Caltech, Stanford, Alabama, Arizona): 
first results [3].
\item The Krasnoyarsk underground (600 mwe) site: a possible two detector
experiment "Kr2Det" to search for very small mixing angles in the 
atmospheric mass parameter region [4]. 
\end{itemize}
Other experiments at Krasnoyarsk are presented here by Yu. Kozlov.
All oscillation experiments are based on the reaction ${\widetilde {\nu}}_{e} + p \rightarrow e^{+} + n$, 
and use $e^{+}$,n delayed coincidence technic. First two of them make use of 
Gd loaded liquid scintillator, for Krasnoyarsk no Gd doping is planned.

\section{The CHOOZ EXPERIMENT}

The CHOOZ detector used a 5 ton target (Fig.1). It was located in an 
underground laboratory (300 mwe) at a distance of about 1 km from two
PWR type reactors of total (nominal) power 8.5 GW(th).

A summary of the data taking from April 1997 and July 1998 is shown 
in the Table 1.

\begin{table*}[hbt]
\setlength{\tabcolsep}{1.5pc}
\newlength{\digitwidth} \settowidth{\digitwidth}{\rm 0}
\catcode`?=\active \def?{\kern\digitwidth}
\caption{Data taking cycles:}
\label{tab:data}
\begin{tabular*}{\textwidth}{@{}l@{\extracolsep{\fill}}cc}
\hline
\\
\vspace{5pt}
            & Time (d) & $\bar W$ (GW) \\
\hline

 Reactor 1 ON    & $85.7$ & $4.03$ \\
 Reactor 2 ON    & $49.5$ & $3.48$ \\
Reactor 1 and 2 ON & $64.3$ & $5.72$ \\
Reactor 1 and 2 OFF & $142.5$ &  \\
\hline
\end{tabular*}
\end{table*}

To identify the ${\widetilde {\nu}}_{e}$ absorption in the target the following selection 
criteria were used: (i) a time delay between the $e^{+}$ and the neutron: 
2 - 100 ${\mu}s$; (ii) spatial conditions: distance from PMT surface of $e^{+}$ and 
neutron candidates $d > 0.3 m$, distance between $e^{+}$ and $n < 1 m$; 
(iii) energy window for n candidates (6 - 12) MeV, and for $e^{+}$  
candidates $\sim$ (1.3 - 8) MeV. Under these conditions the ${\widetilde {\nu}}_{e}$ detection 
efficiency was found to be ${\epsilon} = (69.8 \pm 1.1)\%$ 

Total about 2500 neutrino events were detected during the data 
acquisi-tion periods. The measured neutrino detection rate is 
$2.5 (d^{-1}GW^{-1})$, the neutrino detection rate to background ratio is 10:1 
(typically). Measured positron energy spectrum is in good agreement with 
expected spectrum in no-oscil-lation case (Fig.2).
The ratio $R_{meas/calc}$ of the measured to calculated for no oscillation 
case neutrino detection rates is found to be \\
\vspace{9pt}
$ R_{meas/calc} = 1.01 \pm 2.8 \%(stat) + 2.7 \%(syst)$ \\
\vspace{9pt}
Components of the combined systematic error are listed in the Tab. 2.

\begin{table*}
\setlength{\tabcolsep}{1.5pc}
\caption{Systematic uncertainties}
\label{tab:uncert}
\begin{tabular*}{\textwidth}{@{}l@{\extracolsep{\fill}}c}
\hline
 parameter      &    error \\
\hline
 reaction cross section & $1.9\%$ \\
 number of protons      & $0.8\%$ \\
 detection efficiency   & $1.5\%$ \\
 reactor power          & $0.7\%$ \\
\hline
 combined               & $2.7\%$ \\
\hline
\end{tabular*}
\end{table*}

The CHOOZ'99 oscillation limits are derived by comparing all the 
experi-mental information available to expected no-oscillation values 
and directly depend on the correct determination of the absolute 
value of the ${\widetilde {\nu}}_{e}$ flux, their energy spectrum, the nuclear fuel 
burn up effects and on other issues listed in Table 2. 

It can be seen (Fig.3) that new oscillation constraints are two times 
more stringent than those of the CHOOZ'97. We note also that CHOOZ 
experiment does NOT observe ${\widetilde {\nu}}_{e}$ oscillations in the mass region 
${\Delta}m^{2}_{atm}$ where muon neutrinos oscillate intensively [5].

\section{THE PALO VERDE EXPERIMENT}

The detector is a matrix of $6\times 11$ acrylic cells each 9 m long filled 
with liquid scintillator (12 tons total weigh). The detector is 
installed in an underground (32 mwe) laboratory at a distance of $\sim$ 800 m 
from three reactors whose total power is 11 GW(th). The signature for 
${\widetilde {\nu}}_{e}$ is a fast triple $e^{+}{\gamma}{\gamma}$ coincidence 
followed by a delayed signal from the neutron. The resultant efficiency of neutrino 
detection was found to be ca $16\%$. 

Presently (September, 1999) results of the first 72 days of data 
taking are available. The 3 reactors ON minus 2 reactors ON data give 
neutrino detection rate of $6.4 \pm 1.4$ per day. This number is compatible 
with expecta-tions for no oscillations. The relevant exclusion plot is 
shown in Fig.3. The experiment is continuing and large increase of the 
sensitivity to the mixing angle is expected.

\section{POSSIBLE EXPERIMENT AT KRASNOYARSK}

\subsection{Contra and pro}

- Shall we apply new efforts and search with higher sensitivity for 
the oscillations of reactor ${\widetilde {\nu}}_{e}$ in atmospheric neutrino mass region?
Often the answer is: - Well, may be not now... After all we know 
already from CHOOZ that ${\widetilde {\nu}}_{e}$ can contribute no more than $10\%$ to the 
oscillations of atmospheric neutrinos...

The support comes from theorists [6]: - While atmospheric problems are 
also important, the reconstruction of the neutrino mixing is one of the 
most fundamental problems of the particle physics. In this particular 
case the quantity $sin^{2}2{\theta}$" is directly expressed trough the element $U_{e}$   
of the neutrino mixing matrix 
$sin^{2}2{\theta} = 4U^{2}_{e3}(1 - U^{2}_{e3})$ \quad
($U_{e3}$ is the contribution of the mass-3 state to the electron neutrino 
state: ${\nu}_{e} = U_{e1}{\nu}_{1} + U_{e2}{\nu}_{2} + U_{e3}{\nu}_{3}$). 

\subsection{An idea of the Kr2Det experiment}

The practical goal of the Kr2Det is to decrease, relative to the 
CHOOZ'99, the statistic and systematic errors as much as possible. 
Therefore we turn to the time honored idea of a near detector and 
consider two identical liquid scintillation spectrometers stationed 
underground (600 mwe) at 1100 m $\sim$ 250 m from the reactor. We increase 
the neutrino target masses up to 50 tons and choose a miniature version 
of the KamLAND detector design (KamLAND has a 1000 ton target).

\begin{table*}
\setlength{\tabcolsep}{1.5pc}
\caption{ Expected ${\widetilde {\nu}}_{e}$ detection rates and backgrounds.}
\label{tab:expect}
\begin{tabular*}{\textwidth}{@{}l@{\extracolsep{\fill}}ccc}
\hline
 Parameter   & Distance (m)&  ${\widetilde {\nu}}_{e} (d^{-1}$) & BKG ($d^{-1}$) \\
\hline
 Detector 1 & 1100 & 50 & 5 \\
 Detector 2 &  250 & 900 & 5 \\
\hline
\end{tabular*}
\end{table*}

In no-oscillation case the ratio of the two simultaneously measured 
posi-tron spectra is energy independent. Small deviations from the 
constant value are searched for oscillations. The results are 
independent of the exact knowledge of the ${\widetilde {\nu}}_{e}$ flux and energy 
spectrum, burn up and the reaction cross section, the numbers of the target 
protons... However the difference of the detector energy  scales 
(response functions) should be controlled. By special intercalibration 
procedures we plan to monitor the difference and to introduce necessary
corrections.

Expected $90\%$ CL oscillation limits are shown in Fig.3. It was 
assumed that 40 000 ${\widetilde {\nu}}_{e}$ are detected in 1100 m detector and the
spectrometer scale difference is controlled down to $0.5\%$.

\section{CONCLUSIONS}

The one km baseline experiments at reactors, CHOOZ and Palo Verde,
have successfully reached the atmospheric neutrino mass region and have 
explored there a large area on oscillation parameters plane. This area 
can be considerably expanded and new information on the neutrino 
intrinsic properties obtained.

\section*{Acknowledgements}

I would like thank Professor Froissart for hospitality. I greatly 
appreciate valuable discussions with S. Bilenky, E. Lisi and A. Smirnov. 
Large part of this work was done in collaboration with V. Sinev. This 
study is supported by RFBR.


\begin{thebibliography}{9}
\bibitem{Appo1} M. Appolonio et al. hep-ex/9907037.
\bibitem{Appo2} M. Appolonio et al. Phys. Lett. B420 (1998) 397; hep-ex/9711002.
\bibitem{Boem}  F. Bohem, nucl-ex/9906010.
\bibitem{Mikae} L. Mikaelyan, V. Sinev hep-ex/9908047.
\bibitem{Schol} K.Scholberg, Super-Kamiokande Coll., hep-ex/9905016; 
  M.Nakahata, these Proceedings.
\bibitem{Bilen} S.M. Bilenky, C.Giunti, Phys. Lett. B444 (1998) 379; G.L. Fogli, these 
   Proceedings; A. Smirnov, in WEIN'98, p 180.
\end{thebibliography}
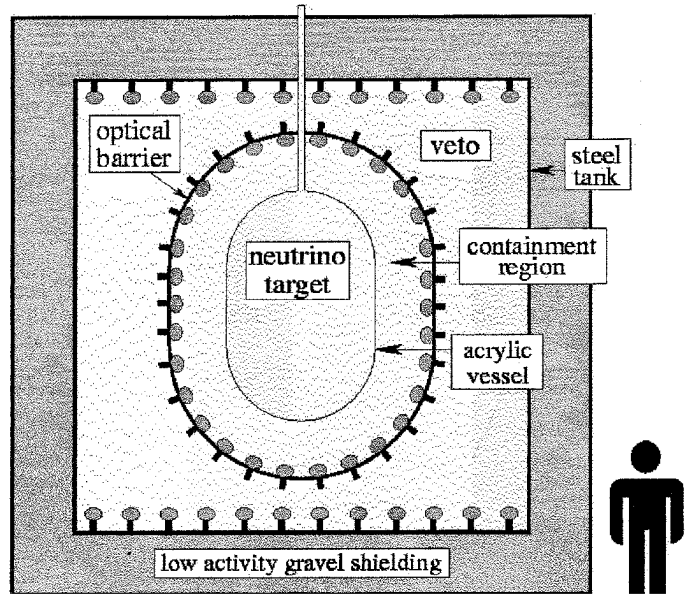
\end{document}